\documentclass[intlimits,twoside,a4paper]{article}

\usepackage{amsmath,amssymb}
\usepackage{graphicx}
\usepackage{wrapfig}

\usepackage[T2A]{fontenc}
\usepackage[cp1251]{inputenc}

\usepackage[eqsecnum]{cmpj2}

\newcommand{\eqb}{\begin{equation}\label{eq:\arabic{section}-\arabic{equation}}}
\newcommand{\eqe}{\end{equation}}

\issue{2013}{16}{1}{13003}

\doinumber{10.5488/CMP.16.13003}

\title[Observation of triple correlations in fluids]%
{Experimental observation of triple correlations \\ in fluids}
\author[M.Ya.~Sushko]{M.Ya.~Sushko\thanks{E-mail:
{mrs@onu.edu.ua}}}
\address{Mechnikov National University, Department of Theoretical
Physics,\\ 2 Dvoryanska St., 65026 Odesa, Ukraine}

\date{Received July 3, 2012, in final form October 29, 2012}
\authorcopyright{M.Ya.~Sushko, 2013}

\begin{document}

\maketitle

\begin{abstract}
We present arguments for the hypothesis that under some
conditions,  triple correlations of density fluctuations in fluids
can be detected experimentally by the method of molecular
spectroscopy. These correlations manifest themselves in the form
of the so-called 1.5- (i.e., sesquialteral) scattering. The latter is of
most significance in the pre-asymptotic vicinity of the critical
point and can be registered along certain thermodynamic paths. Its
presence in the overall scattering pattern is demonstrated by our
processing experimental data for the depolarization
factor. Some consequences of these results are discussed.
\keywords  density fluctuations, critical opalescence,
1.5-scattering, depolarization factor
\pacs 05.40, 05.70.Fh, 05.70.Jk, 64.70.Fx, 78.35.+c
\end{abstract}

\section{Introduction}

The intensity $I$ of light scattered by a one-component fluid
drastically increases  as the critical point is approached
\cite{Fab65,Cum74}. The physical nature of this critical
opalescence phenomenon is well-known, i.e., an increase in the magnitude
of permittivity (in fact~--- density) fluctuations and the development of long-range
correlations between them. The result is that $I$ is contributed
to by not only single scattering effects, but also by those  of
higher multiplicities. One could expect that $I$ provides certain
information about higher-order correlation functions for the
density fluctuations.

However, a common view is that it is only double scattering~\cite{Oxt74,Cha74,Adz75,Kuz75,Boo75,Cha83} together, probably, with Andreev's scattering (due to fluctuations in the
distribution function of thermal fluctu\-a\-tions)~\cite{And74,Ale1,Mal1} and triple scattering~\cite{Tra77} that
contribute significantly near the critical point. The pertinent
theories are based upon  quasi-Gaussian statistics for the
fluctuations. As a consequence, any information on the
irreducible parts of higher-order correlation functions is lost
and multiple scattering is viewed as parasitic.

By contrast, we pursue the idea that the so-called 1.5-
(sesquialteral) scattering, caused by triple correlations of
density fluctuations, contributes significantly to $I$. In this
report, we support this statement by  our results of processing
extensive experimental data~\cite{Tra80} on the depolarization
factor near the critical point. In our view, these results
strongly indicate that the 1.5-scattering is noticeably present,
under certain conditions, in the overall scattering pattern.
Moreover, the idea of 1.5-scattering allows us to give simple
explanations for the anomalies in the behavior of $I^{-1}$ in the
gravity field~\cite{Ale79,Ale2} and those of the Landau-Plazcek
ratio near the $\lambda$-line~\cite{Win73,Con75,Vin78}, which
were  observed long ago, but have been interpreted
controversially.

It is also important to note that, according to our own theory and
in view of earlier estimates~\cite{Cha74}, from among different
three-point configurations of density fluctuations $\delta \rho$
(located at points ${\bf r}$, ${\bf r}_1$, and ${\bf r}'$), the
major contribution to the integrated 1.5-scattering intensity
$I_{1.5}$ is made by those in which two successively scattering
fluctuations merge on the macroscopic scale ($| {\bf r} - {\bf
r}_1 | \to 0$; technically, this is a consequence of the
replacement of the internal electromagnetic field propagators in
the iterative series for the scattered field with their leading
short-range singularities). As a result, $I_{1.5}$ is determined
by the Fourier transform of the pair correlation function
$\langle \left[ \delta \rho (\rm {\bf r}) \right]^{2} \delta
\rho ({\bf r}')  \rangle$. Provided Polyakov's hypothesis
\cite{Pol70} (see also~\cite{Pat82,Car87}) of conformal symmetry
of critical fluctuations is valid, the latter is expected to
vanish at the critical point (see appendix~A). It follows  that
the recovery of the 1.5-scattering contribution from $I$ and
a scrutinized study of its behavior along appropriate thermodynamic
paths ending up at the critical point provide a unique
opportunity for experimental verification of Polyakov's hypothesis
\cite{Pol70} for systems with scalar order parameters.

\section{Theory of 1.5-scattering}

\subsection{General expression}

The theory of  1.5-scattering was proposed in~\cite{Sush1} and
developed further in~\cite{Sush2,Sush3}; some additional numerical
estimates were made in~\cite{Sush4,Sush5}. We assume that
molecular light scattering from condensed matter is a result of
re-emission of light not only by single fluctuations, but also
by compact groups of fluctuations. By compact we understand any group
of fluctuations all the distances between which are much shorter
than the wavelength $\lambda$ of the probing light in the medium.
Physically, scattering by such a group is single. The overall
polarized single-scattering spectrum is, therefore, given by the
series~\cite{Sush2}
\begin{equation} \label{a1}
 I({\rm {\bf q}},\omega ) = {\sum\limits_{n,m \geqslant 1}
{I_{nm} ({\rm {\bf q}},\Omega )}} ,
\end{equation}
\noindent where
\begin{equation}\label{a2}
I_{nm} ({\rm {\bf q}},\Omega ) \propto \,\left( { -
{\frac{{1}}{{3\varepsilon _{0}}} }} \right)^{n + m - 2} \times
{\frac{{1}}{{\pi }}}\Re{\int\limits_{0}^{ + \infty} {{\rm
d}t}} {\int\limits_{V} {\rm d}{\rm {\bf r}}{\left\langle {\left[
{\delta \varepsilon ({\rm {\bf r}},t)} \right]^{n}\left[ {\delta
\varepsilon (0,0)} \right]^{m}} \right\rangle }\,{\rm e}^{{\rm
i}\Omega t - \ri{\rm {\bf q}} \cdot {\rm {\bf r}}}}
\end{equation}
is the contribution from a pair of compact groups of $n$
and $m$ permittivity fluctuations [attributed further to density
fluctuations, $\delta \varepsilon \approx \left(\partial
\varepsilon /\partial \rho \right)_T \delta \rho$],
$\varepsilon_0$ is the equilibrium value of the permittivity,
$\Omega $ and ${\rm {\bf q}}$ are the changes in the light
frequency and wavevector due to scattering, and the scattering
volume $V$ is included into the proportionality coefficient.

It is only the term $I_{11}({\rm {\bf q}},\Omega)$ in equation
(\ref{a1}) that has been associated so far with the single
scattering. The 1.5-scattering intensity is defined as $I_{1.5}
({\rm {\bf q}},\Omega)= I_{12} ({\rm {\bf q}},\Omega) + I_{21}
({\rm {\bf q}},\Omega)$.

\subsection{Hydrodynamic region, $qr_\mathrm{c}\ll 1$}\label{ss1}
Far enough from the critical point, where the correlation radius
$r_\mathrm{c} \ll \lambda$ and nonlocal correlations between fluctuations
can be ignored, the integrated 1.5-scattering intensity can be
expressed in terms of the third moment of thermodynamic
density fluctuations $\Delta\rho$~\cite{Sush1}:
\begin{equation}\label{a3}
I_{1.5} \propto -\frac{2}{3 \varepsilon_0 } \left(\frac{\partial
\varepsilon}{\partial \rho} \right)^3 \widetilde{V}\langle(\Delta
\rho )^3\rangle = -\frac{2}{3 \varepsilon_0} \left(\rho
\frac{\partial \varepsilon}{\partial \rho} \right)^3
\frac{k^{2}_{\mathrm{B}}T^{2}}{\widetilde{V}} \left[ 2
\beta_T^{2}+\left(\frac {\partial\beta_T}{\partial P}
\right)_{T,V}\right],
\end{equation}
where $\beta_T$ is the isothermal compressibility of the fluid and
$\widetilde{V} $ is a macroscopic volume over which the
fluctuations $\delta \rho $ are averaged to single out their
thermodynamic parts $\Delta\rho$. We suggest that $\widetilde{V} $
is slightly dependent on temperature far away from the critical
point, but  $\widetilde{V} \propto r_\mathrm{c}^3 \propto \beta_T^{3/2}$ in
the critical region.

Calculations with the van der Waals and Dieterici equations of
states give the estimates
\[ I_{1.5}  \propto - \frac{1}{\varepsilon_{0}}\left(\rho
\frac{\partial\varepsilon} {\partial\rho} \right)^{3}
\frac{k^{2}_{\mathrm{B}}T^{2}}{\widetilde{V}} \cdot 6 P_{c}\omega
\beta_T^{3}
 \]
and
\[ I_{1.5} \propto - \frac{2}{3\varepsilon_{0}}\left(\rho
\frac{\partial\varepsilon} {\partial\rho} \right)^{3}
\frac{k^{2}_{\mathrm{B}}T^{2}}{\widetilde{V}} \cdot
\left(\beta_T^2 +4  P_\mathrm{c}\omega \beta_T^{3}\right),
 \]
respectively, where $\omega\equiv\rho_\mathrm{c}/\rho - 1$, $|\omega| \ll 1$ is the
deviation of $\rho$ from the critical value $\rho_\mathrm{c}$ and $P_\mathrm{c}$
is the critical pressure. It follows that the 1.5-scattering can
become  of significance in those domains in the
$(\tau,\omega)$-plane where $\omega \neq 0$, but $\beta_T$ is
sufficiently large. Then, $I_{1.5} \propto \omega \beta_T^{3/2}$
for a non-critical isochore, but $I_{1.5} \propto \beta_T^{1/2}$
or even $I_{1.5} \to 0 $ for the critical one.

A distinctive feature of the 1.5-scattering contribution is that
it is not positive definite: for instance, $I_{1.5} <0$ in the
region where $\omega >0$ and $\tau>0$, at least.

\subsection{Fluctuation region, $qr_\mathrm{c}\gg 1$}\label{ss2}

Understanding, in this section,  $\rho$ as a scalar order
parameter, we see that formulas (\ref{a1}) and (\ref{a2}) agree
with the hypothesis of algebra of fluctuating quantities
\cite{Pat82}. Then, within the first order of $\epsilon$-expansion
and in the long-wave limit $q \to 0$, the critical index of
$I_{1.5}$, defined by  $I_{1.5} \propto |\tau| ^{-\mu_{21}}$, is
estimated to be $\mu_{21}\approx 0.67$ for $\omega =0$
\cite{Sush1}. This value is close to an earlier estimate of 0.7
given in~\cite{Pat82}. Correspondingly, $I_{1.5} \propto
\beta_T^{1/2} $ on the critical isochore and in the immediate
vicinity of the critical point.

This result can be refined using the algebra of fluctuating
quantities (see appendix~A). However, it is more important to
emphasize that it disregards the conformal invariance hypothesis~\cite{Pol70}. If the latter is indeed valid, then the
orthogonality relation holds for fluctuating quantities with
different scaling dimensions (see~\cite{Pat82,Car87}), that is,
$I_{1.5} \to 0$ as the critical point is approached.

\subsection{Intermediate  region, $qr_\mathrm{c}\lesssim 1$}
This region is of special interest to us because it is typical of
actual experiment. Taking into account that correlations between
fluctuations $\delta \rho$ remain relatively weak, we argue
\cite{Sush2} that the convolution-type approximation  \[
{\left\langle {\rho _{{\rm {\bf k}}_{1}}  \rho _{{\rm {\bf
k}}_{2}}  \rho _{{\rm {\bf k}}_{3}}} \right\rangle} \approx -
{\frac{{2{c}'}}{{k_{B} T\sqrt {V}}} }G(k_{1} )G(k_{2} )G(k_{3}
)\,\delta _{{\rm {\bf k}}_{3} , - {\rm {\bf k}}_{1} - {\rm {\bf
k}}_{2}} \] can be used for the three-point correlation function
of density fluctuations. Here, $\rho _{{\rm {\bf k}}} $ is the
Fourier component of $\delta \rho ({\rm {\bf r}})$,  $G(k) \equiv
\langle {|\rho _{{\rm {\bf k}}}|^2 \rangle} $, and ${c}'$ is a
$\bf k$-independent function of temperature and density.
Calculations with the Ornstein-Zernike expression for $G(k)$ then give:
\begin{equation}\label{a4}
I_{1.5}(q) \propto \frac{c'}{3 \pi \varepsilon_0 } \left(\rho
\frac{\partial \varepsilon}{\partial \rho} \right)^3 \frac{ \rho^3
k^{2}_{\mathrm{B}}T^{2} \beta_T^3} {qr_\mathrm{c}^4\left(1+q^2r_\mathrm{c}^2
\right)}\arctan\left(\frac{qr_\mathrm{c}}{2}\right).
\end{equation}
Requiring that in the limit $qr_\mathrm{c}\ll 1$, equation (\ref{a4})
transforms into equation (\ref{a3}), we recover ${c}'$ through the
third thermodynamic moment of density fluctuations, with an
accuracy to a positive proportionality constant:
\begin{equation}\label{a5}
c' \propto - \left[ 2 \beta_T^{2}+\left(\frac
{\partial\beta_T}{\partial P} \right)_{T,V}\right]
\left(\rho\beta_T\right)^{-3}.
\end{equation}

Extrapolation of formulas (\ref{a4}) and (\ref{a5})  on the
fluctuation region shows that (see appendix~B) $c' \to 0$ and,
therefore, $I_{1.5}\to 0$ as both $\tau \to 0$ and $\omega \to 0$,
which is in accordance with the conformal invariance hypothesis.

The structure of the 1.5-scattering spectrum in the intermediate
region is discussed in~\cite{Sush3}.

\section{Depolarization factor}

\subsection{Theoretical considerations}

Now, we are in a position to scrutinize the effect of
1.5-scattering on the depolarization factor $\Delta$ as a function
of temperature (in fact, $\beta_T$) and the geometrical size $L
\sim V^{1/3}$ (volume $V$) of the scattering system. Suppose that
the following contributions to $I$ are present in the intermediate
region $qr \lesssim 1$: (1) the ``standard'' intensity $I_{11}
\propto V \beta_T$ of polarized single scattering due to density
fluctuations~\cite{Fab65,Cum74}; (2) the intensity $I_{1.5}$ of
polarized 1.5-scattering~\cite{Sush1,Sush2,Sush3}; (3) the
intensity $I_{\rm a}\propto V$ of depolarized single scattering
due to anisotropy fluctuations~\cite{Fab65} (it is virtually
insensitive to the critical point); (4) the intensity $I_{2\rm p}
\propto V^{4/3} \beta_T^2$ of polarized double scattering due to
density fluctuations~\cite{Oxt74,Cha74,Adz75,Kuz75,Boo75,Cha83};
(5) the intensity $I_{2\rm d} \propto V^{4/3}\beta_T^2$ of
depolarized double scattering due to density fluctuations
\cite{Oxt74,Cha74,Adz75,Kuz75,Boo75,Cha83}; (6) the intensity
$I_{\rm A}\propto V \beta_T^{1/2}$ of depolarized single
scattering due to fluctuations in the distribution function of
thermal fluctuations (Andreev's scattering)~\cite{And74}. Then,
$\Delta$ is given by
\begin{equation} \label{c1} \Delta= \frac{I_{\rm a}+
I_{2\rm d}+I_{\rm A}}{I_1 +I_{1.5}+ I_{2\rm p} }\,.
\end{equation}

In view of the individual temperature dependences of the above
contributions and under the condition $I_{1.5}=0$, $\Delta$ as a
function of $\beta_T$ is expected to decrease first, then reach a
minimum, and then increase again. Such a behavior, indeed observed
in the experiment, is considered as a manifestation of double
scattering effects. However, as we show later, the presence of the
1.5-scattering contribution does not alter this qualitative
behavior of $\Delta$ as a function of $\beta_T$.

Thus, expression (\ref{c1}) should be transformed in order to obtain an
experimentally-measurable function whose behavior significantly
depends on whether the 1.5-scattering contributes to $\Delta$ or
not~\cite{Sush1}. Rewriting (\ref{c1}) as
\begin{equation} \label{c2}
\frac{I_{2\rm d}}{I_1\Delta}=\frac{1 +I_{1.5}I_1^{-1}+ I_{2\rm
p}I_1^{-1}} {1+ I_{\rm a}I_{2\rm d}^{-1}+ I_{\rm A}I_{2\rm
d}^{-1}}
\end{equation}
and taking into account the specific features of the intensity
contributions, we immediately arrive at the relation
\begin{equation} \label{c3}
\frac{L\beta_T}{\Delta}\propto\frac{1+a\beta_T^{1/2}+bL\beta_T}{1+cL^{-1}
\beta_T^{-2}+dL^{-1}\beta_T^{-3/2}}
\end{equation}
valid for the intermediate region $qr\lesssim 1$. Here,  the
coefficients $b$, $c$, and $d$ are practically
temperature-independent and positive constants. The coefficient $a
\propto c' \arctan\left(\frac{qr_\mathrm{c}}{2}\right)/r_\mathrm{c}$ is due to the
1.5-scattering contribution and is not positive definite. If the
1.5-scattering is negligible, then $a=0$ and the right-hand side
in formula (\ref{c3}) is a monotonous increasing function of
$\beta_T$. With the 1.5-scattering present, this monotonous behavior
is expected to be violated. The effect should be most pronounced
in the following two cases.

(1) The critical point is approached along a noncritical isochore
$\omega >0$. Then, $I_{1.5} \propto -\beta_T^{3/2}$ and $a$ is
close to a negative constant.

(2) The critical point is approached along the path where $\tau
\to 0$, $\omega\to 0$, and  $I_{1.5}>0$. The relative magnitude
$I_{1.5}/I_1$ of the 1.5-scattering should start decreasing
somewhere  due to the temperature law $I_{1.5}
\propto \beta_T^{1/2}$ coming into play~\cite{Pat82,Sush1} (see section~\ref{ss2})
or as a consequence of the conformal invariance~\cite{Pol70,Pat82}. As such a path, the liquid branch of the
coexistence curve can be quoted.

Thus, by varying the temperature ($\beta_T$) and density
($\omega$) of the scattering system, we hope to ``stick out''  the
1.5-scattering contribution from among the others. It should
manifest itself as a non-monotonous behavior of the
experimentally-measurable quantity $L\beta_T/\Delta^{-1}$ with
$\beta_T$. The fact that the scattering contributions involved
depend differently on $L$, provides an additional powerful option
for analysis. Some results obtained by processing the extensive
depolarization factor data~\cite{Tra80} are presented in figures
\ref{fig-Fig1}--\ref{fig-Fig14}. They generalize our earlier
results~\cite{Sush4}.

\subsection{Data processing}

\subsubsection{Noncritical isochores $\rho < \rho_\mathrm{c}$, $\tau >0$}

\begin{figure}[!t]
\centerline{
\includegraphics[width=0.49\textwidth]{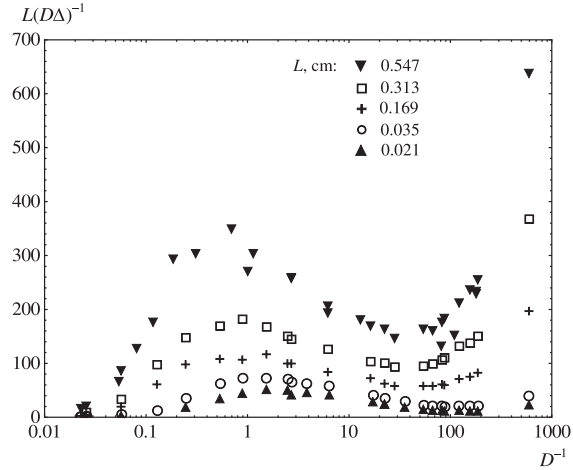}
}
\caption{$L(D\Delta)^{-1}$ versus $D^{-1}$ along the $\omega=
6.8\times 10^{-3}$ isochore of xenon for five values of $L$, based
on experimental data~\cite{Tra80}. From left to right, $\tau$
decreases from $7.8 \times 10^{-2}$ to $1.2 \times 10^{-5}$. Three
segments can be distinguished on each of these log-linear plots. \label{fig-Fig1}}
\end{figure}

Figure \ref{fig-Fig1} represents the log-linear plots of the
quantity $L(D\Delta)^{-1} \propto L\beta_T\Delta^{-1}$ as a
function of the parameter ($k_0$ is the wavevector in vacuum of
the incident light)
\[ D^{-1}= \frac{k_0^{4}}{144 \pi^2}
\left(\varepsilon_0 -1 \right)^{2}\left(\varepsilon_0 +2
\right)^{2} k_{\rm B}T\beta_T \propto \beta_T
\] for xenon along the $\omega= 6.8 \times 10^{-3}$ isochore and
five values of $L$ (in cm). The parameter $ D $ (in m) is a
convenient measure of the distance  to the critical point~\cite{Tra79}.
It is evaluated in~\cite{Tra80} as a function of
temperature and density by using the Clausius-Mossotti relation
for $\varepsilon_0$ and the restricted linear model equation of
state~\cite{Sen78} for $\beta_T$. These calculations are claimed
to be most reliable for the region not very close to and not far
away from the critical point, i.e., the one  of special interest to
us.

\begin{figure}[!b]
\centerline{
\includegraphics[width=0.49\textwidth]{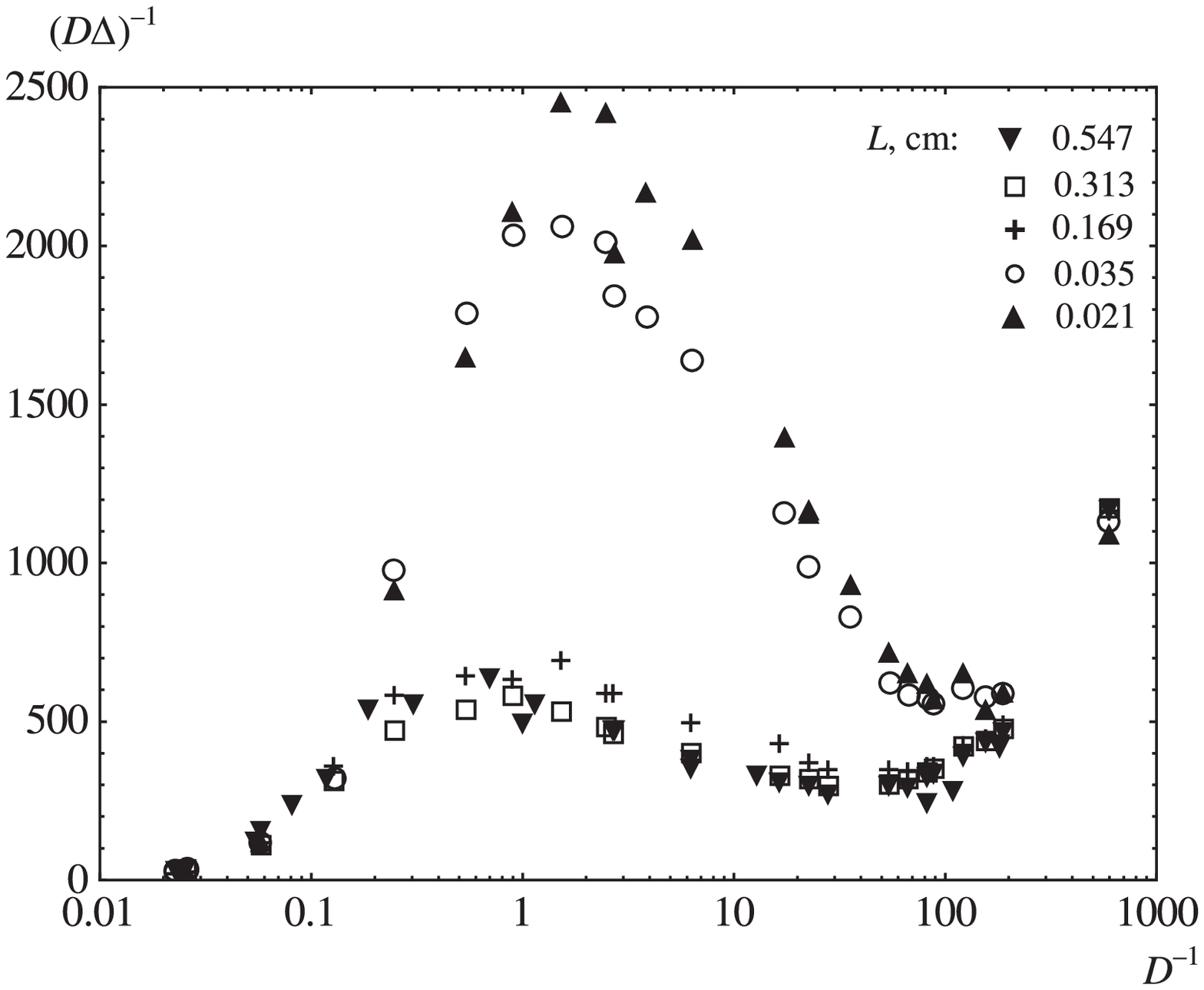}
\hfill%
\includegraphics[width=0.485\textwidth]{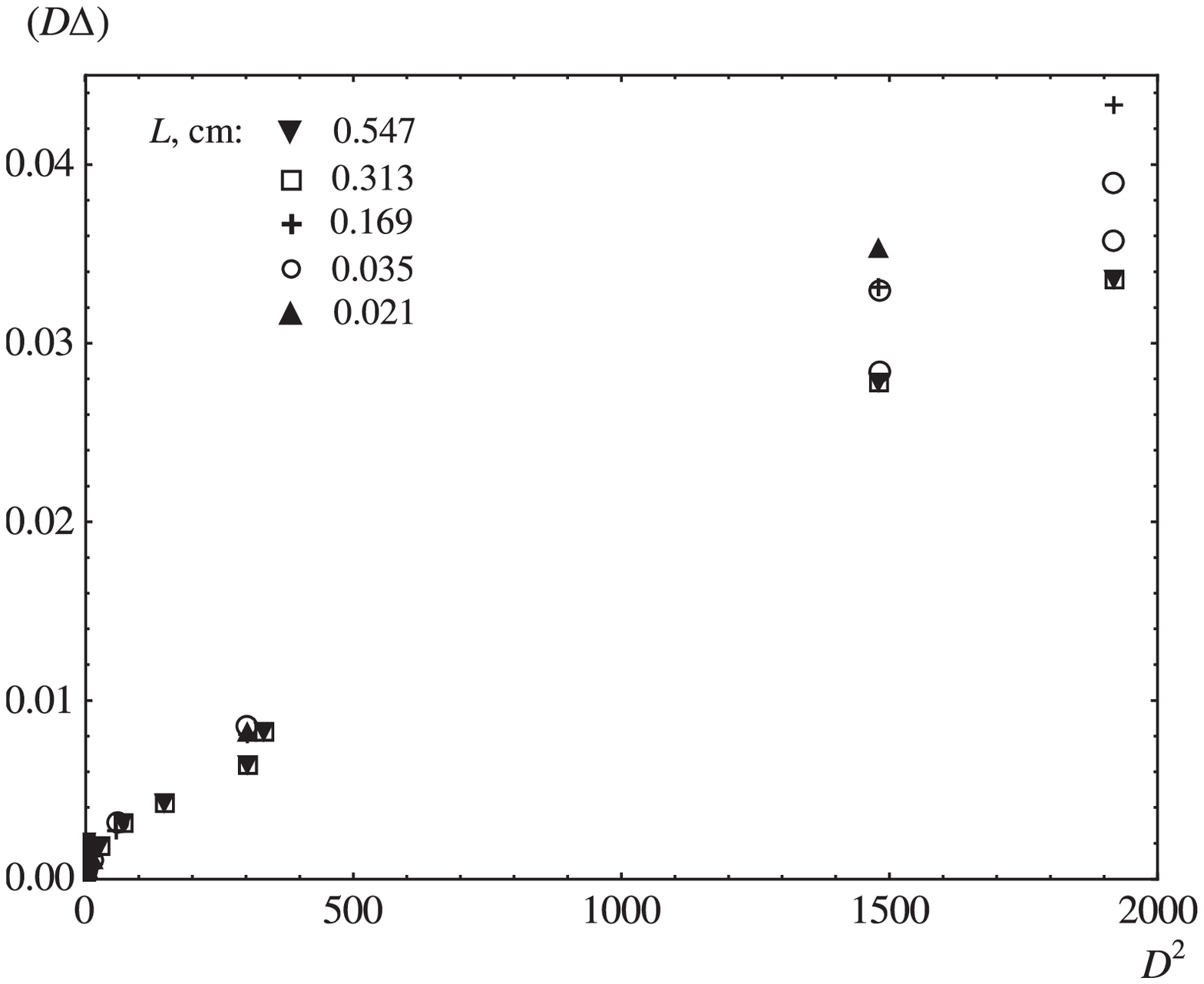}
}
\parbox[t]{0.5\textwidth}{%
\caption{$(D\Delta)^{-1}$ versus $D^{-1}$ for the data shown in
figure~\ref{fig-Fig1}.\label{fig-Fig2}}}%
\parbox[t]{0.5\textwidth}{%
\caption{$(D\Delta)$ versus $D^{2}$ for  segments A. From right to
left, $\tau$ decreases from $7.8 \times 10^{-2}$ to $3.3 \times
10^{-3}$.  \label{fig-Fig3}}
}%
\end{figure}

As $\tau$  reduces from $7.8 \times 10^{-2}$ to $1.2 \times
10^{-5}$, three typical temperature intervals are clearly seen in
figure~\ref{fig-Fig1}. We shall refer to them as segments  A
(leftmost), B (central) and C (rightmost). It is easy to note that
simple division of $L(D\Delta)^{-1} $ by $L$ and changing to
$(D\Delta)^{-1}$ as a function of $D^{-1}$ for different values of
$L$  transforms the original plots dissimilarly on these segments:
the plots tend to merge on A and C, but invert the vertical
ordering and disperse on B (figure~\ref{fig-Fig2}). This fact
implies that $(D\Delta)^{-1}$ is contributed to by terms with
differing functional dependences on $L$. Our further analysis of
them is guided by relation (\ref{c3}).

Suppose that on segments A, i.e., the most distant from the critical
point, $I_1$  prevails  much over  $I_{1.5}$ and $I_{2\rm p}$.
Then, relation (\ref{c3}) takes the form
\[ \frac{L\beta_T}{\Delta}\propto\frac{1}{1+cL^{-1}
\beta_T^{-2}+dL^{-1}\beta_T^{-3/2}}\,.
\]
It follows that the
dependence of $(D\Delta)$ upon $D^2$ should be close to a linear
one, with the slope independent of $L$ and, if the Andreev
contribution is noticeable, a slight concavity: $(D\Delta)
\propto{\rm const} + cD^2 + d(D^2)^{3/4}.$
Figure~\ref{fig-Fig3} does exhibit, at least approximately, such
kind of dependence. The study of the latter could, in principle,
provide experimental estimates for the magnitude of $I_{\rm A}$.
However, the discussion of this issue is beyond the scope of
the present report.

\begin{figure}[!t]
\centerline{
\includegraphics[width=0.49\textwidth]{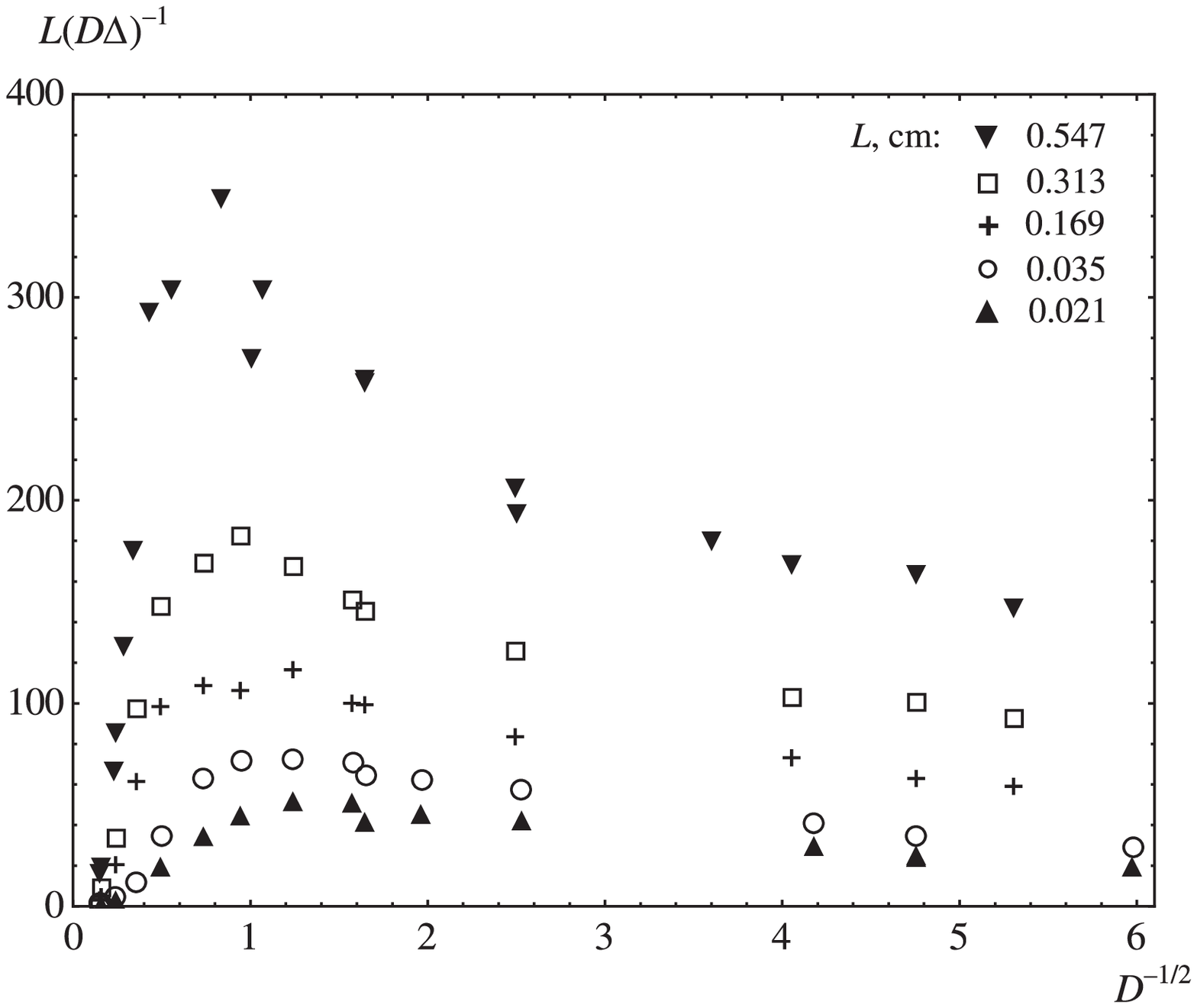}
\hfill%
\includegraphics[width=0.49\textwidth]{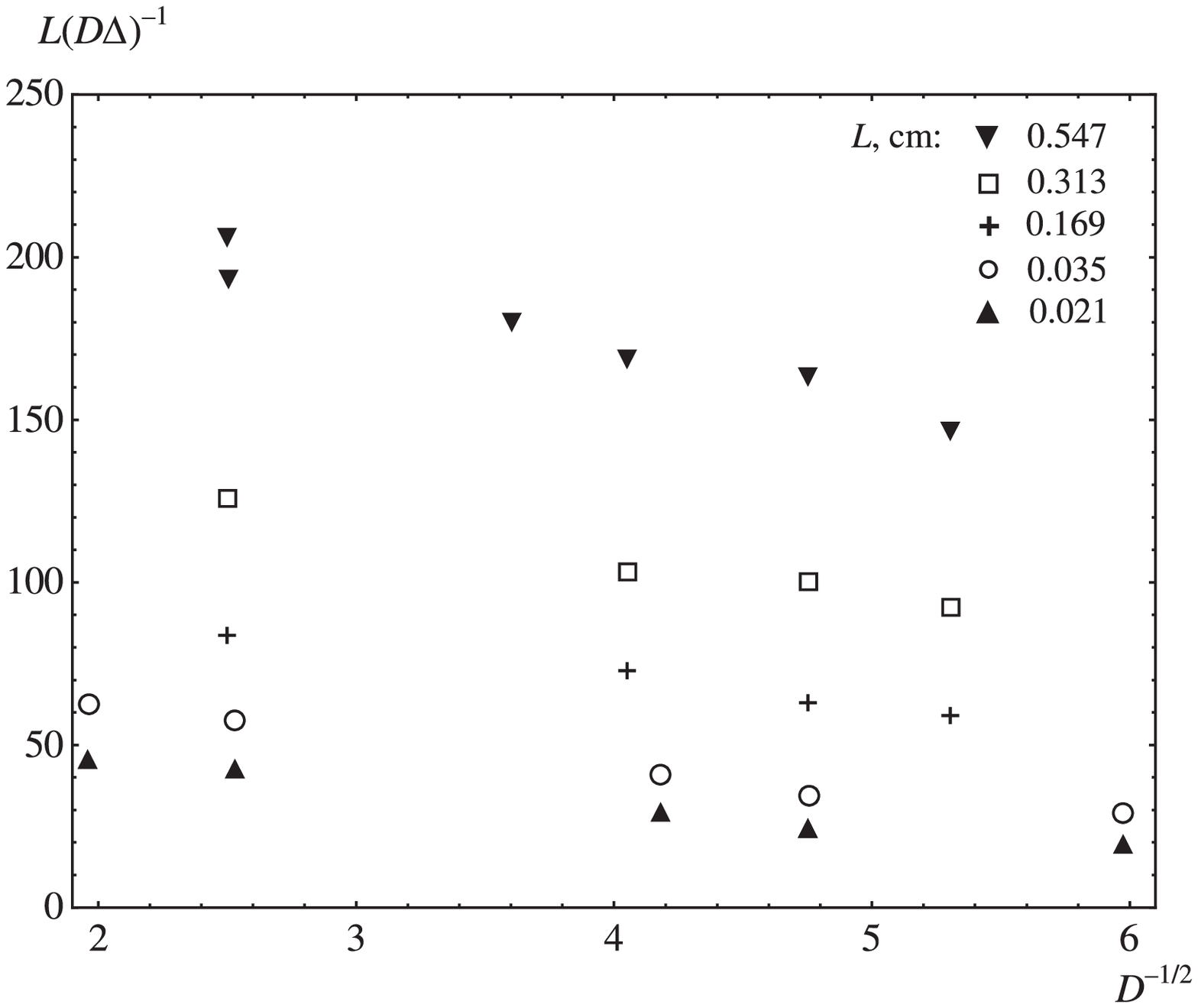}
}
\vspace{-5mm}
\parbox[t]{0.5\textwidth}{%
\caption{$L(D\Delta)^{-1}$ versus $D^{-1/2}$ for segments  A and
B. From left to right, $\tau$ decreases from $7.8 \times 10^{-2}$
to $1.4 \times 10^{-4}$.} \label{fig-Fig4}
}%
\parbox[t]{0.5\textwidth}{%
\caption{$L(D\Delta)^{-1}$ versus $D^{-1/2}$ for the portions of
segments B where $\tau$ decreases from $9.8 \times 10^{-4}$ to
$1.4 \times 10^{-4}$.} \label{fig-Fig4a}
}%
\end{figure}

\begin{figure}[!b]
\centerline{
\includegraphics[width=0.50\textwidth]{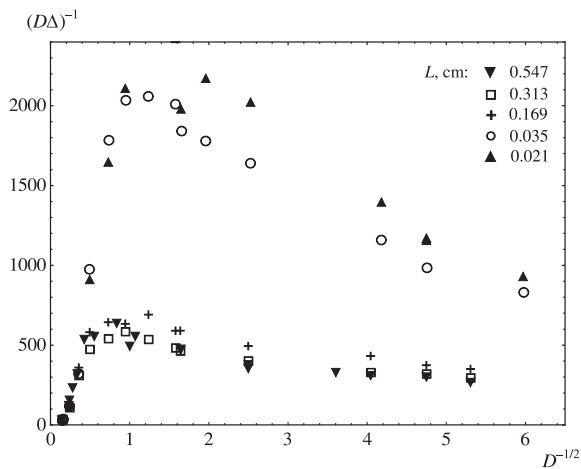}
}
\caption{$(D\Delta)^{-1}$ versus $D^{-1/2}$ for the data shown in
figure~\ref{fig-Fig4}.} \label{fig-Fig5}
\end{figure}

On segments B, where we expect $I_{2 \rm d}$ to dominate over
$I_{\rm a}$ and $I_{\rm A}$, but $I_{2 \rm p}$ to remain
relatively weak as compared to $I_{1}$  and $I_{1.5}$, relation
(\ref{c3}) takes the form
\[ \frac{L\beta_T}{\Delta}\propto {1 +a \beta_T^{1/2}}.\]
At $\omega > 0$, $L(D\Delta)^{-1}$ should decrease with $D^{-1/2}$
by the linear law $L(D\Delta)^{-1} \propto - D^{-1/2}$, with
negative and equal slopes for different values of $L$
(figures~\ref{fig-Fig4} and \ref{fig-Fig4a}). Correspondingly, the
dependences of $(D\Delta)^{-1}$ upon $- D^{-1/2}$ should be linear
on B, with slopes inversely proportional to $L$, but, as was
already mentioned, merge on A (figure \ref{fig-Fig5}).

\begin{figure}[!t]
\centerline{
\includegraphics[width=0.49\textwidth]{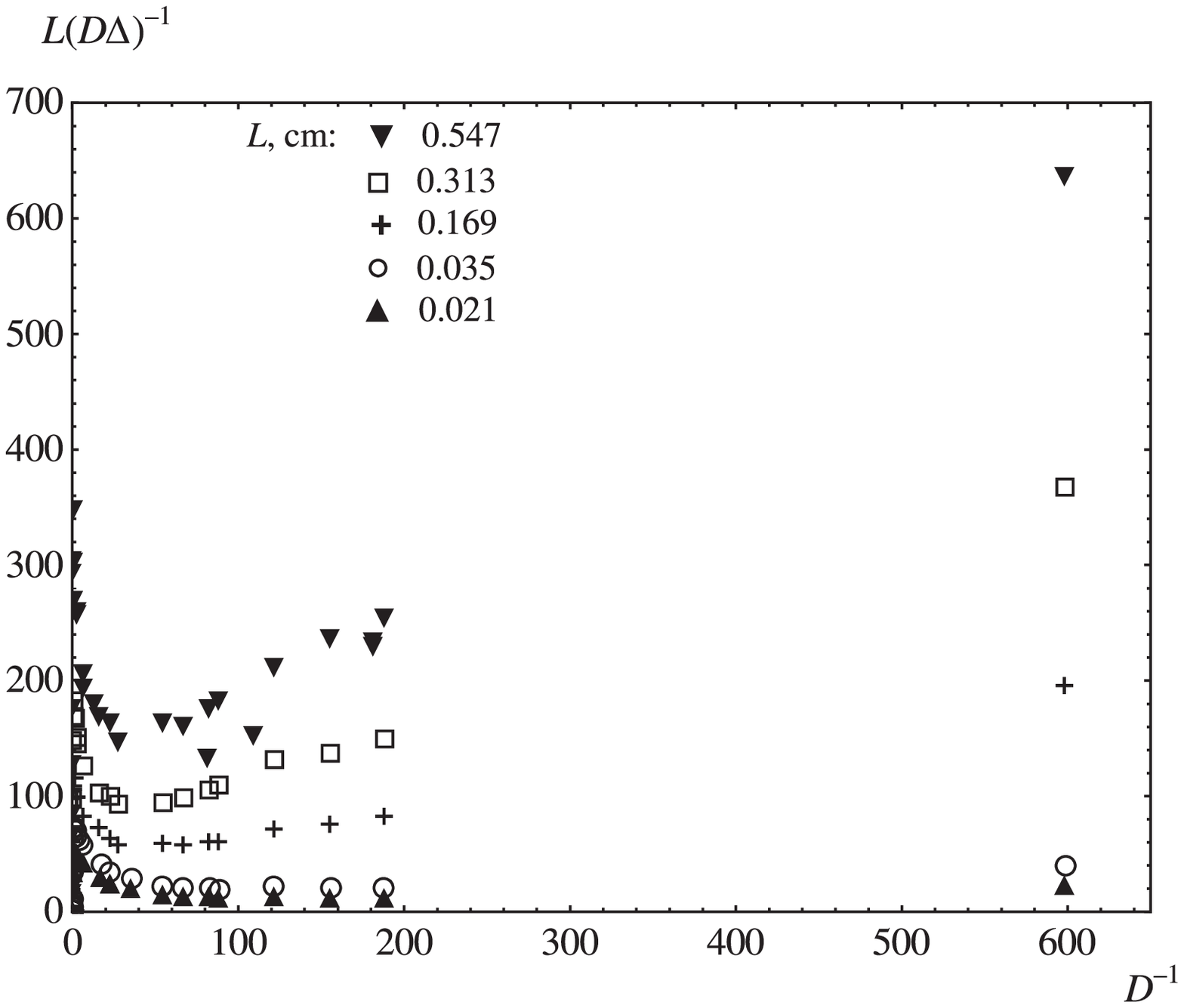}
\hfill%
\includegraphics[width=0.49\textwidth]{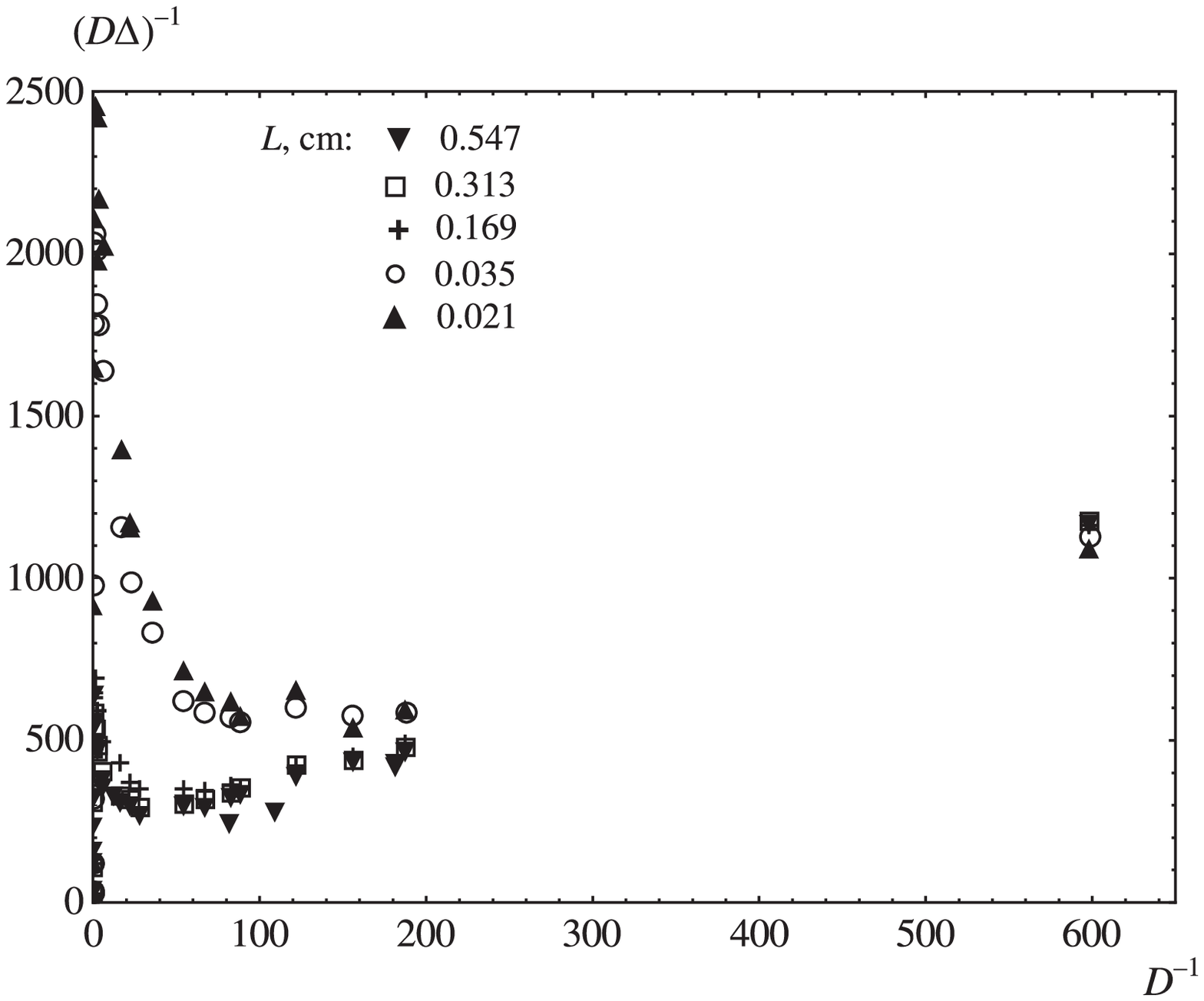}
}
\parbox[t]{0.5\textwidth}{%
\caption{$L(D\Delta)^{-1}$ versus $D^{-1}$ for the entire $\omega=
6.8\times 10^{-3}$ data plotted with a linear $D^{-1}$ scale.
Segments C stand out as $\tau $ decreases rightwards from $8.6
\times 10^{-5}$ to $1.2 \times 10^{-5}$.} \label{fig-Fig6}
}%
\parbox[t]{0.5\textwidth}{%
\caption{$(D\Delta)^{-1}$ versus $D^{-1}$ for the entire $\omega=
6.8\times 10^{-3}$ data plotted with a linear $D^{-1}$ scale.}
\label{fig-Fig7}
}%
\end{figure}

Finally, segments C are formed  mainly by single and true
double scatterings. Relation (\ref{c3}) takes the form
\[\frac{L\beta_T}{\Delta}\propto 1+ bL \beta_T
\]
and we expect $L(D\Delta)^{-1}$ to increase with $D^{-1}$ by a linear
law, with slope proportional to $L$ (figure \ref{fig-Fig6}). The
$(D\Delta)^{-1}$ versus $D^{-1}$ plots  for different values of
$L$ should approach a single straight segment as $D^{-1}$
increases (figure~\ref{fig-Fig7}).

\begin{figure}[!b]
\centerline{
\includegraphics[width=0.50\textwidth]{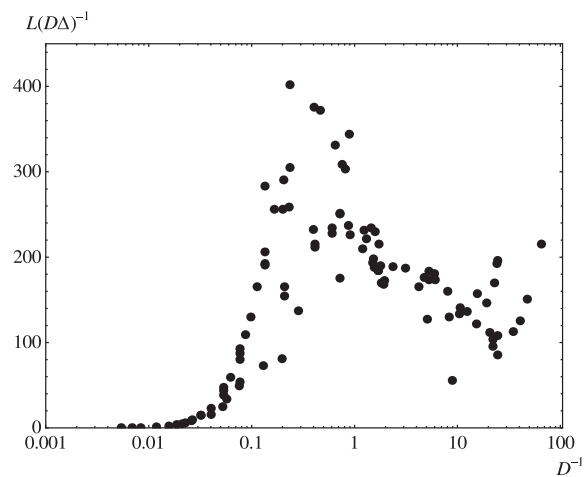}
}
\caption{$L(D\Delta)^{-1}$ versus $D^{-1}$ along the liquid branch
of the coexistence curve of xenon;  $L = 0.547\, {\rm cm}$. From
left to right, $\tau$ changes from  $- 9.2\times 10^{-2}$ to
$-2.9\times 10^{-5}$.} \label{fig-Fig11}
\end{figure}

%\subsection{Noncritical isochore $\rho > \rho_\mathrm{c}$, $\tau >0$}
\subsubsection{Liquid branch of the coexistence curve}

The dependence of $L(D\Delta)^{-1}$ upon $D^{-1}$ along the liquid
branch of the coexistence curve of xenon is shown in figure~\ref{fig-Fig11}. It agrees well with our expectations.

Thus, the above processing of experimental data~\cite{Tra80}
clearly reveals the presence in  the overall scattering pattern of
a contribution which we associate with the 1.5- (sesquialteral)
molecular light scattering.

\subsection{Numerical estimates}

Now, we  present quantitative estimates of the magnitude of
1.5-scattering intensity. They were  obtained by fitting the
$L(D\Delta)^{-1}$ versus $D^{-1}$ data for the entire $\omega=
6.8\times 10^{-3}$ isochore and then used to reproduce the
original $\Delta$ versus $D$ data~\cite{Tra80}.

In view of formulas (\ref{c2}) and (\ref{c3}), the fitting
function was taken in the form
\[    f(x) = \frac{x^2}{C+x^2}\left(K + A x^{1/2} +B x  \right),
\qquad x \equiv D^{-1}, \]
and the following two sets of
coefficients were chosen: $K_1 = 581.066$, $A_1 = -63.6968$, $B_1
= 3.66569$, $C_1 = 0.012 $ and $K_2 = 607.508$, $A_2 = -89.0932$,
$B_2 = 5.79532$, $C_2 = 0.012$. The relative magnitudes $r_{1.5}
\equiv {I_{1.5}}/{I_{1}}$ and $r_{2 {\rm p}} \equiv {I_{2 {\rm
p}}}/{I_{1}}$ of the 1.5-scattering and polarized double
scattering, as compared to the single scattering, were estimated
as
\[ r_{1.5}  = \frac{A x^{1/2}}{K}\,,
\qquad  r_{2 {\rm p}}  = \frac{B x}{K}\,.\]
To calculate $\Delta $
with $f_i$, our theoretical estimate~\cite{Mal1} $I_{2\rm
d}\approx \frac{1}{8} I_{2\rm p}$ was additionally used. Then,
\[
\Delta = \frac{Bx}{8 K f(x)}\,.
\]

\begin{figure}[!h]
\centerline{
\includegraphics[width=0.49\textwidth]{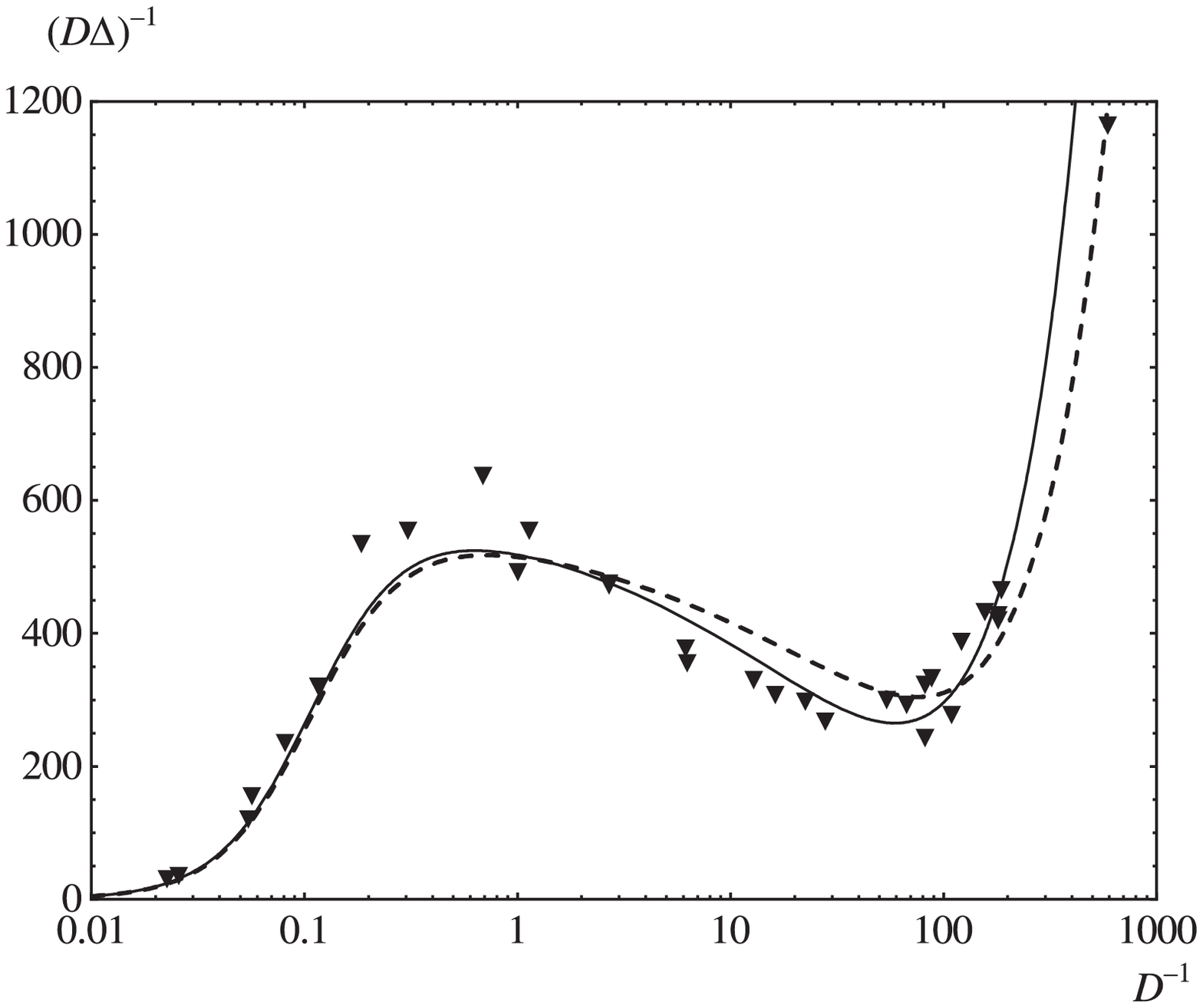}
\hfill%
\includegraphics[width=0.49\textwidth]{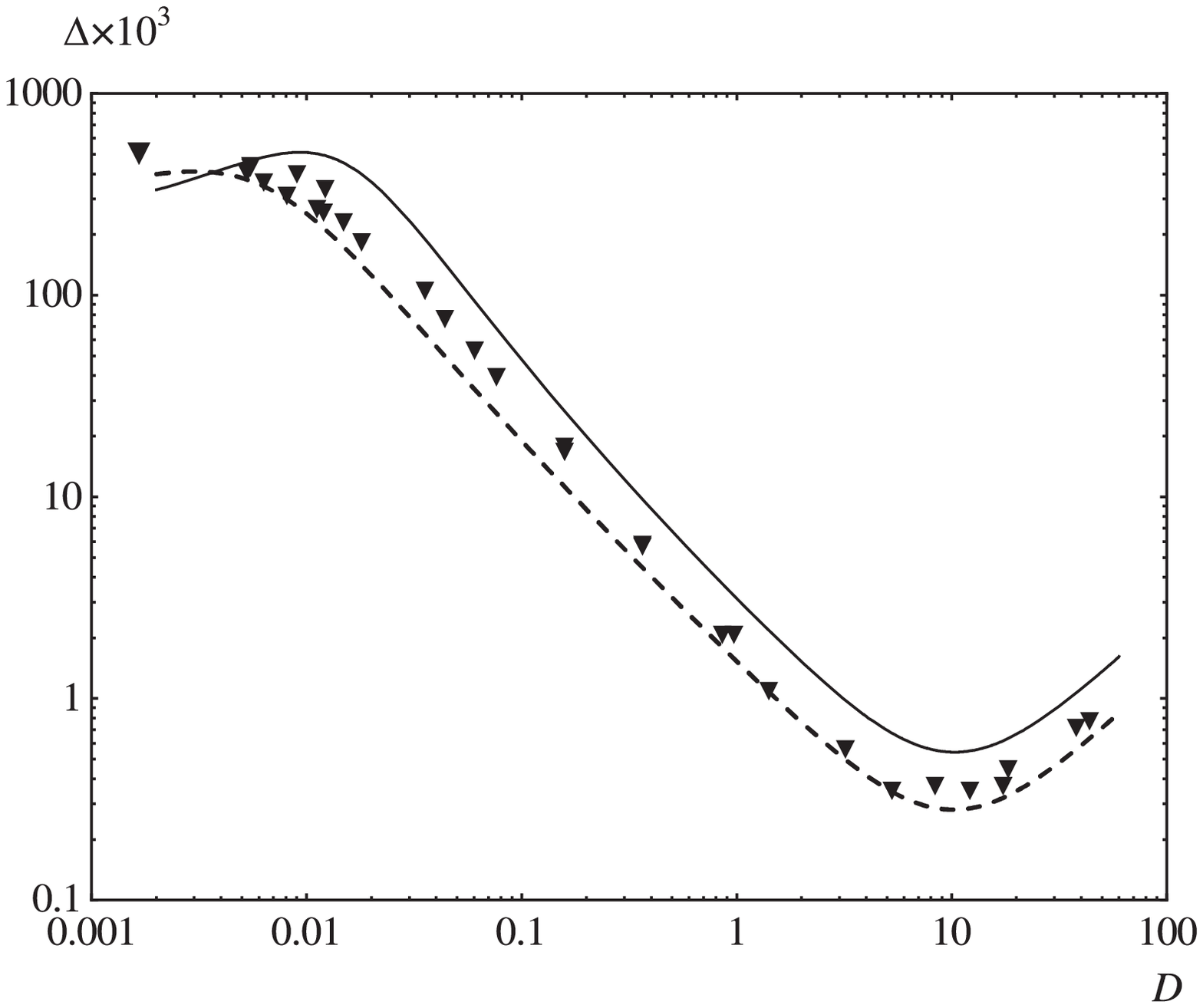}
}
\vspace{-3mm}
\parbox[t]{0.5\textwidth}{%
\caption{Fitting the $(D\Delta)^{-1}$ versus $D^{-1}$ data
($\omega= 6.8\times 10^{-3}$, $L = 0.547\, {\rm cm}$) with $f_1$
(dashed line) and $f_2$ (solid line).}\label{fig-Fig12}
}%
\parbox[t]{0.5\textwidth}{%
\caption{Fitting the $\Delta$ versus $D$ data ($\omega= 6.8\times
10^{-3}$, $L = 0.547\, {\rm cm}$) with $f_1$ (dashed line) and
$f_2$ (solid line).}\label{fig-Fig13}
}%
\end{figure}

\begin{figure}[!t]
\centerline{
\includegraphics[width=0.50\textwidth]{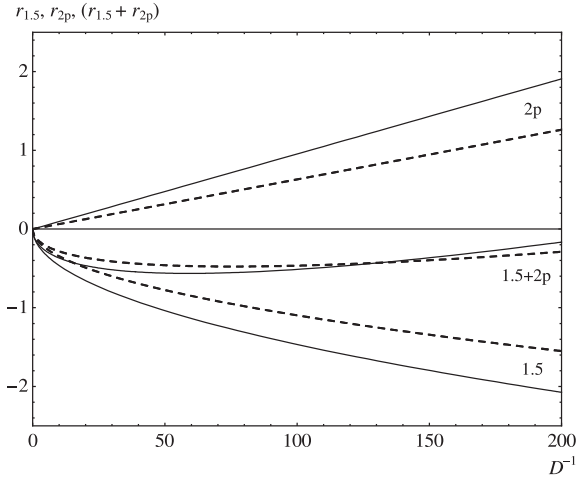}
}
\caption{$I_{1.5}/I_{1}$,  $I_{2 {\rm{p}}}/I_{1}$, and $\left(
I_{1.5} + I_{2 {\rm p}}\right) /I_{1}$ along the $\omega=
6.8\times 10^{-3}$ isochore of xenon, $L = 0.547 \,{\rm cm}$, as
estimated with $f_1$ (dashed line) and $f_2$ (solid
line).}\label{fig-Fig14}
\end{figure}

The results are demonstrated by figures~\ref{fig-Fig12}--\ref{fig-Fig14}. They clearly show that in the
intermediate region, the intensities $I_{1.5}$ and $I_{2 {\rm p}}$
reach magnitudes comparable with that of $I_1$, but are opposite
in sign and tend to compensate for each other. These facts are
surprising. They contradict the common view that multiple
scattering contributions come into play gradually as the critical
point is approached. In other words, they imply an asymptotic
nature of the iterative series for the overall scattering
intensity near the critical point. They can also be interpreted in
the sense that triple and quadruple correlations in fluids
contribute, at least to light scattering effects, in opposite
directions.

The agreement of our fitting results with the $\Delta $-data~\cite{Tra80} (figure \ref{fig-Fig13}) is  also unexpectedly good.

\section{Conclusion}
The above estimates  have prompted us to identify other
experiments where the situation is favorable for 1.5-scattering to
come into play. First of all, of interest are the studies on light
scattering from critical fluids  under the earth's gravity. Due to
the gravity effect, the system is spatially in\-ho\-mo\-ge\-neous
in the vertical direction. A negative 1.5-contribution is expected
to appear in the light scattered from the fluid layers located
above the level of critical density. For such a layer, the total
scattering intensity $I$ should,  as a function of temperature,
start decreasing somewhere as the critical point is approached. In
other words, the $I^{-1}$ versus $\tau$  dependence should have a
minimum at some $\tau_0$. The effect was indeed registered, for
instance, in freon 113~\cite{Ale79,Ale2}. Our estimations for the
minimum location, $\tau_0 \sim 10^{-3}$ for heights up to $20\,
{\rm mm}$ ($\tau >0$), agree well with experiment.

We suggest that a specially-designed processing of the
gravity-induced height- and temperature dependences of $I$,
obtained for systems with a scalar order parameter, is a feasible
opportunity for singling out the 1.5-scattering contribution and
verifying Polyakov’s conformal invariance hypothesis.

To finish, we mention that the studies of the spectral
distribution in critical opalescence spectra are of great interest
as well. In particular,  we have proved that in the presence of
1.5- and double scattering effects, the ratio of the integrated
intensities of the Rayleigh and Brillouin components takes the
form
\begin{equation}
\label{d1} R_{{\rm exp}} =  R \frac{1 + a_{1.5} r_{1.5}+ a_{2{\rm
p}} r_{{2{\rm p}}}} {1 + b_{1.5} r_{1.5}+ b_{2{\rm p}} r_{2{\rm
p}}}\,,
\end{equation}
where $R= \gamma -1$ is the well-known Landau-Placzek ratio~\cite{Lan82} for single scattering ($\gamma \equiv c_P/c_V $,
$c_P$ and $c_V$ being the specific heats at constant pressure and
volumes). The coefficients $a_{1.5}$ and $b_{1.5}$ are given in
\cite{Sush3}, whereas $a_{2{\rm p}}$ and $b_{2{\rm p}}$ can be
recovered from the results~\cite{Cha77}:
\begin{equation}
\label{d11} a_{2{\rm p}} = 1-\frac{3}{2\gamma}+\frac{1}{2( \gamma
-1)}\, ,\qquad
 b_{2{\rm p}}=2 - \frac{3}{2\gamma}\,.
\end{equation} Suggesting that $r_{1.5} =0$, it is not
difficult to verify, based on experimental data~\cite{Ath73}  for
${\rm He}^4$, that the double scattering alone should cause
$R_{{\rm exp}}$ to exceed $R$  as the $\lambda$-line is approached
along a high-pressure isobar. Such a fact was indeed registered
\cite{Vin78}, but in most other experiments in this series the
tendency was direct opposite~\cite{Win73,Con75,Vin78}. We
attribute the reduction in $R_{{\rm exp}}$ to the effect of
1.5-scattering.

Our detailed calculations of the above effects will be presented
elsewhere.

\clearpage

\section*{Appendix A}

Let $A_k({\bf r})$ be a complete set (algebra) of fluctuating
scalar quantities with scaling dimensions $\Delta_k$: under the
scaling transformations ${\bf r} \rightarrow\lambda {\bf r}$,
$A_k(\lambda {\bf r}) \rightarrow \lambda^{-\Delta_k} A_k({\bf
r})$. According to the local algebra hypothesis (see, for
instance,~\cite{Pat82}), the scalar function $\delta\rho ({\bf
r})$ and, therefore, the scalar function $\left[\delta\rho ({\bf
r})\right]^{2}$ can be developed into the series
\[
\delta\rho ({\bf r}) = \sum_{k=1}^\infty a_k A_k({\bf r}), \qquad
 \left[\delta\rho ({\bf r})\right]^2 =
\sum_{m,n=1}^\infty b_{mn} B_{mn}({\bf r}), \] where the
coefficients $b_{mn}$ are simply related to the coefficients
$a_k$, and $B_{mn}({\bf r}) = A_m({\bf r}) A_n({\bf r})$ are
scalar quantities with definite scaling dimensions $\Delta_{mn}=
\Delta_{m}+\Delta_{n}$: $B_{mn}(\lambda {\bf r}) \rightarrow
\lambda^{-\Delta_{mn}} B_{mn}({\bf r})$. Correspondingly, the pair
correlation function $\langle \delta \rho ({\bf r}')\left[
\delta \rho (\rm {\bf r}) \right]^{2} \rangle$ is given by
a linear combination of pair correlators \linebreak $K_{k,mn}(|{\bf r}'-{\bf
r}|)\equiv \langle A_k({\bf r}')B_{mn}({\bf r})\rangle$.

For a $d$-dimensional system with scalar order parameter, only the
Fourier transforms of those $K_{k,mn}$ can reveal a singular
behavior  near the critical point which satisfy the condition
$\Delta_k +\Delta_{mn}< d$. In the first-order
$\epsilon$-expansion $\Delta_k=
\frac{1}{2}k(2-\epsilon)+\frac{1}{6}k(k-1)\epsilon$, where
$\epsilon=4-d$~\cite{Pat82}. If $d=3$, then the relevant
correlators are $K_{1,11}$, $K_{1,12}=K_{1,12}$, and $K_{2,11}$,
each involving two scalar quantities with different scaling
dimensions. Once Polyakov's conformal symmetry hypothesis holds
true and the system is spatially homogeneous and isotropic, they
all vanish at the critical point due to the orthogonality relation
(see~\cite{Pat82,Car87} and the literature cited therein).

\section*{Appendix B}

As one of the ways for evaluating $c'$ in the fluctuation region,
we can use the asymptotic equation of state~\cite{Sush6}
\[
\pi (\tau ,\omega ) = M\tau + {\frac{{1}}{{2\Gamma _{0}}} }
\frac{1}{(1+\omega)^2} \tau \,{\left| {\tau} \right|}^{\gamma - 1}
- D_{0} \omega {\left| {\omega } \right|}^{\delta - 1},\] which
immediately  follows from the requirements that it leads to (a)
the correct asymptotic behavior of   a limited number of the
fluid parameters along the selected thermodynamic paths (the
susceptibility $\chi =\rho^2 \beta_T$ along the critical isochore,
the critical isotherm equation, and the derivative of pressure
with respect to temperature at the critical point) and (b) reveal
a Van-der-Waals-type loop below the critical point. In this
equation, $\pi = {{P} \mathord{\left/ {\vphantom {{P} {P_\mathrm{c}}} }
\right. \kern-\nulldelimiterspace} {P_\mathrm{c}}}  - 1$, $\tau = {{T}
\mathord{\left/ {\vphantom {{T} {T_\mathrm{c}}} } \right.
\kern-\nulldelimiterspace} {T_\mathrm{c}}}  - 1$, $\omega = {{V}
\mathord{\left/ {\vphantom {{V} {V_\mathrm{c}}} } \right.
\kern-\nulldelimiterspace} {V_\mathrm{c}}}  - 1$, $\Gamma_0$ and $D_0$
are the critical amplitudes for $\chi$ and the critical isotherm,
respectively, and $M$ is a constant. The definition
\[ \beta _{T} = - \frac{1}{V}\left(\frac {\partial V}{\partial P}
\right)_{T,V}= - \frac{1} {P_\mathrm{c}(1+\omega)} \left( \frac{\partial
\omega} {\partial \pi}  \right)_{\tau},
\]
relations
\[\left(\frac {\partial\beta_T}{\partial P}
\right)_{T,V} =\left(\frac {\partial\beta_T}{\partial P}
\right)_{T,N}=\left(\frac {\partial\beta_T}{\partial V}
\right)_{T,N}\left(\frac {\partial V}{\partial P}
\right)_{T,N}=-(1+\omega) \beta_T \left(\frac
{\partial\beta_T}{\partial \omega} \right)_{\tau},\] and formula
(\ref{a5}) then yield
\[ c' \propto -P_\mathrm{c} D_0 \delta (\delta -1)
\rho^{-3} \omega |\omega|^{\delta -3}.
\]

%\clearpage

\newpage
\ukrainianpart

\title{Експериментальне спостереження потрійних кореляцій \\ у рідинах}
\author{М.Я. Сушко}
\address{Одеський національний університет імені І.І. Мечникова, вул. Дворянська, 2, 65026 Одеса, Україна}

\makeukrtitle

\begin{abstract}
\tolerance=3000%
Наведено аргументи на користь гіпотези, що при певних умовах
методом молекулярної спектроскопії можна зареєструвати потрійні
кореляції флуктуацій густини в рідинах. Ці кореляції проявляють
себе у вигляді так званого 1.5- (тобто півторакратного) розсіяння,
яке є найбільш суттєвим в перед\-асимптотич\-ній області критичної
точки та може бути зареєстроване вздовж певних термодинамічних
шляхів. Його присутність у загальній картині розсіяння
демонструється результатами обробки відомих експериментальних
даних для коефіцієнта деполяризації. Обговорено деякі наслідки
цих результатів.
\keywords  флуктуації густини, критична опалесценція, розсіяння
кратності 1.5, коефіцієнт деполяризації
\end{abstract}

\end{document}